# In Praise and in Search of Highly-Polarizable Semiconductors

R. Jaramillo[1], Jayakanth Ravichandran[2,3]

1. Department of Materials Science and Engineering, Massachusetts Institute of Technology, Cambridge, MA 02138, USA
2. Mork Family Department of Chemical Engineering and Materials Science, University of Southern California, Los Angeles, CA 90089, USA
3. Ming Hsieh Department of Electrical and Computer Engineering, University of Southern California, Los Angeles, CA 90089, USA

The dielectric response of materials underpins electronics and photonics. At high frequencies, dielectric polarizability sets the scale for optical density and absorption. At low frequencies, dielectric polarizability determines the band diagram of junctions and devices, and nonlinear effects enable tunable capacitors and electro-optic modulators. More complicated but no less important is the role of dielectric response in screening bound and mobile charges. These effects control defect charge capture and recombination rates, set the scale for insulator-metal transitions, and mediate interactions among charge carriers and between charge carriers and phonons.

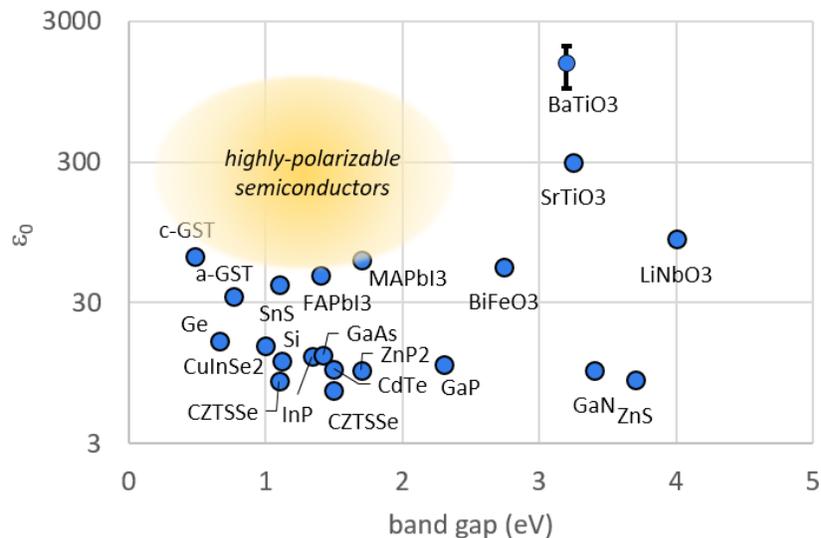

**Figure 1:** Dielectric polarizability *vs.* band gap for semiconductors and complex oxides.[1–9] We use the low frequency dielectric constant ($\varepsilon_0$) as to represent polarizability because it is widely-reported. The two points labeled CZTSSe represent low- and high-band gap $Cu_2ZnSn(S,Se)_4$. c-GST and a-GST represent crystalline and amorphous $Ge_2Sb_2Te_5$. In FAPbI3 and MAPbI3, FA and MA represent formamidinium and methylammonium, respectively. The light orange region labeled "highly-polarizable semiconductors" represents a hypothesis and a suggested research focus.

## In search of highly-polarizable semiconductors

Established semiconductor material platforms are based on a motif of tetrahedral covalent bonding and relatively light elements obeying the octet rule. As a result, these materials have a narrow range of dielectric susceptibility, with low-frequency values ($\varepsilon_0$) on the order of 10 being typical for group-IV, III-V, and II-VI systems (the dielectric constant for many of these semiconductors is enhanced at high-frequency, near the band gap); see **Figure 1**, where we present data for $\varepsilon_0$ and the band gap for a number of well-known materials. Strong and variable dielectric response is associated with more complex crystal structures and heavier elements. Notable examples include perovskite-structured oxides and halides. These systems include the (Ba,Sr)TiO$_3$ system with strong and tunable dielectric susceptibility that is essential for radio frequency (RF) communications, and the paradigmatic MAPbI$_3$ (MA = methylammonium) for which strong dielectric susceptibility and electron-phonon coupling is thought to underpin excellent ambipolar transport properties and solar energy conversion efficiency.[10–12] Another example is the Ge-Sb-Te system in a defect-ordered, rock salt structure which materials feature resonant bonding, large optical density, and phase-change functionality.[13]

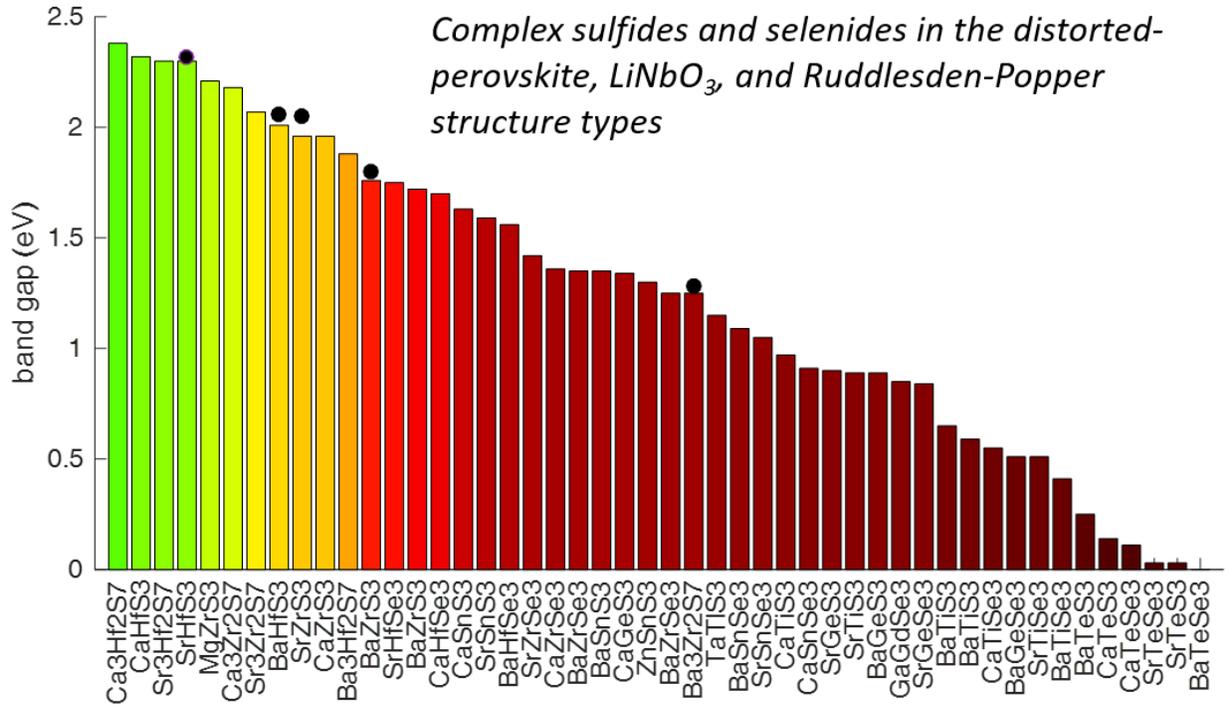

**Figure 2:** Band gap of complex sulfide and selenide materials in the distorted-perovskite, LiNbO$_3$, and Ruddlesden-Popper structure types predicted by theory.[14–20] The colors approximate band gap values in the visible, from Ca$_3$Hf$_2$S$_7$ (green) to CaSnS$_3$ (red). Materials with band gap below the visible (near-IR to metallic) are represented as dark red. Experimental results exist for a few of these materials and are shown as black points.[16,17,21–23]

Complex-structured electronic materials such as oxide perovskites are a rich field of study because they combine characteristics of ionic and covalent bonding in crystal structures that are tolerant of distortions and chemical substitution. The data in **Figure 1** show that the largest dielectric susceptibility is found among complex oxides, and that among semiconductors, large susceptibility is found in materials with complex atomic structures and ionic bonding. The four highly-polarizable complex oxides owe their large dielectric constant to proximity to ferroelectric-paraelectric phase transitions, where the balance between covalent and ionic bonding is pivotal.[24] Ionic characteristics result in strong electron correlation as screening is reduced by the loss of valence electrons. Covalent characteristics result in strong electron-phonon coupling and orbital ordering. These phenomena arise from directional bonding, and illustrate the fact that the electronic and magnetic structures can be are highly-sensitive to bond angles and lengths. The crystal fields, formal oxidation states, and bond geometry can be modified by collective instabilities, chemical substitution, and mechanical strain.[25] This balance of ionic and covalent characteristics in flexible crystal structures gives rise to phenomena such as colossal magnetoresistance, high temperature superconductivity, and ferroelectricity.[25–31] Phenomenological descriptions developed since the 1970s by Goodenough, Zaanan, Sawatzky, Allan, and others send a consistent message: controlling the balance between ionic and covalent bonding is key to manipulating complex-structured electronic materials.[32–36]

**Figure 1** defines a zone (colored yellow) of highly-polarizable semiconductors for which few examples are known. Using the low-frequency dielectric constant as a representative value, we can define highly-polarizable as $\varepsilon_0 \gg 10$; here we loosely-define a semiconductor as a material with a band gap in the infrared (IR) or visible and that can be made conductive through chemical doping or illumination. Based on the underlying trends in band gap and dielectric susceptibility with solid state chemistry, we hypothesize that complex-structured chalcogenide semiconductors will fill this zone of highly-polarizable semiconductors.

How should we search for highly-polarizable semiconductors? A straightforward way is to follow the materials science dictum that properties follow from structure, by substituting chalcogens for oxygen in complex oxide crystal structures. The expectation that complex chalcogenides offer functional properties similar to complex oxides, but with more covalent bonding and smaller band gap, is supported by recent results. $Ae_3B_2S_7$ ($Ae$ = Ca, Sr; $B$ = Zr, Hf), $ZnSnS_3$, and $PbTi(O,S)_3$ are predicted by theory to be ferroelectrics with band gap in the range 1 - 2 eV.[14,15,37,38] The $Ca_3Zr_2S_7$ - $Sr_3Zr_2S_7$ - $Ba_3Zr_2S_7$ system may be functionally similar to the $CaTiO_3$ - $SrTiO_3$ – $BaTiO_3$ (BST) system, with non-polar end-members ($Ba_3Zr_2S_7$, space group $P4_2/mnm$; $Sr_3Zr_2S_7$, space group $P4_2/mnm$; $SrTiO_3$, space group $Pm\bar{3}m$; $CaTiO_3$, space group $Pnma$) and one polar end-member ($Ca_3Zr_2S_7$, space group $A2_1am$ or $Cmc2_1$; $BaTiO_3$, space group $P4mm$) in a binary system with complete solid solubility. For the $PbTi(O,S)_3$ alloy, the decrease of ionic charge and increase in covalent bonding with increasing sulfur content was shown explicitly using electronic structure theory.[37] Experimental results have confirmed the decrease in band gap upon substituting sulfur for oxygen in perovskite-structured materials: from 3.6 to 1.8 eV for $BaZrX_3$, and from 5.5 to 1.9 eV for $CaZrX_3$ ($X$ = O or S).[17,21,39,40] Recent results (by the authors and others) show that the ambipolar opto-electronic properties of complex chalcogenides are promising. A study on green solid state lighting found that $SrHfS_3$ can be doped both $n$- and $p$-type, and has strong band-edge photoluminescence even as cold-pressed ceramics; strong photoluminescence has also been observed in powder samples of $SrZrS_3$ and $BaZrS_3$.[22,23] Work on $Ba_3Zr_2S_7$ found that single-crystals exhibit remarkably slow minority-carrier recombination, suggestive of long minority-carrier lifetime in the bulk and self-passivated surfaces.[16]

Theory is a guide for discovering highly-polarizable semiconductors. In **Figure 2** we present the theoretically-predicted band gap of sulfides and selenides in the perovskite, Ruddlesden-Popper, and $LiNbO_3$ structure types. The data are calculated by density functional theory (DFT) and published elsewhere.[14–20] The band gap values span from the visible to the far-IR, with many that are relevant for solid state lighting, photovoltaics, and photonic communications. Most of the materials presented in **Figure 2** have not been reported synthesized; lacking experimental evidence, there is no guarantee that the phases represented (composition and structure) are thermodynamically stable, or that the predicted band gap values are accurate. However, early experimental results are promising. Five of these materials have been made and their band gap measured, and the results (back points) are quantitatively consistent with the theoretical predictions.

In **Figure 2** we focus on materials without substantial electronic structure anisotropy. Many complex chalcogenides form in more anisotropic crystal structures, such as $BaTiS_3$ in the needle-like phase with face-sharing $TiS_6$ octahedra, and I-V-VI$_2$ chalcogenides built from by polymer-like arsenic-chalcogenide chains. These and many other such materials are interesting and potentially useful in their own right, especially for photonic applications taking advantage of properties such as large birefringence and nonlinearity.[41–43]

*In praise of highly-polarizable semiconductors*

Why focus on dielectric polarizability? Many fundamental and useful advances have come from studying the dielectric properties of electronic materials. Polarization and screening are central to understanding metal-insulator transitions, which are characterized by a diverging screening length. The dielectric response of a crystal lattice to itinerant electrons is described by electron-phonon coupling, and is responsible for phenomena including polarons and superconductivity (it was polaronic transport that inspired Bednorz and Müller to look for superconductivity in complex copper oxides). Strong interaction of phonons with excitons can produce broadband, white light emission, and can support polaron polaritons that are of interest for quantum electronics and lasing.[44–46] Strong nonlinear electric field effects and polar instabilities underlie tunable dielectrics and the electro-optic effect, which are needed for RF and photonic technologies.[10,47] Polar materials also feature giant and unusual light-matter interactions that may become technologically-useful if realized in semiconductors with IR and visible light.[48,49]

Dielectric response may also be an important factor in determining the minority-carrier lifetime of semiconductors. Minority-carrier devices such as light-emitting diodes (LEDs), photovoltaics (PV), photodiodes, and bipolar junction transistors (BJTs) rely on the separation of electron and hole quasi-Fermi levels. This separation defines the free energy of electron-hole pairs and is limited by the rates of electron-hole recombination. Successful technologies are built from materials with low rates of defect-assisted, Shockley-Read-Hall (SRH) minority-carrier recombination.[50–52] The meteoric rise of halide perovskite PV performance has been enabled by minority-carrier lifetime values well above 100 ns and equally impressive values for diffusion length and quantum efficiency.[53,54] These values are comparable to materials such as GaAs and $CuInS_2$ that benefit from decades of continuous research, and underlie the exceptional performance of halide perovskites in PV, LEDs, lasers, and radiation detectors.[55,56] A growing body of research suggests that strong electron-phonon coupling and low-energy, anharmonic polar lattice vibrations act to screen carriers and reduce recombination rates.[12,57–59]

*Outlook*

The above-presented results suggest that the space of complex-structured chalcogenides includes many semiconductors with band gap and optical properties that are useful for opto-electronics. However, the field remains relatively unexplored. The dielectric response and functional properties such as ferroelectricity and the electro-optic effect are largely unknown, as are electronic transport properties. Studies to-date have focused on room-temperature (experimental) and zero-temperature (theoretical) properties, and the effects of changing temperature on the atomic and electronic properties are unknown. For instance, many complex chalcogenides may be incipient ferroelectrics, thermodynamically-adjacent to polar phases, which could be explored by temperature-dependent measurements of thermodynamic response variables including dielectric susceptibility. Phase transitions (*e.g.* ferroelectric, ferro-elastic) and coupled phenomena such as electro-mechanical and electro-optic effects have not been studied. Our recent results showing extremely slow carrier recombination rates in $Ba_3Zr_2S_7$ are promising for PV applications, but are only a first step towards device research, and the scientific connection between these properties and the material dielectric susceptibility has not been explored.[16] We expect that results and interest will multiply as methods are developed to make high-quality samples (single crystals, thin films, nanostructured materials) of materials such as those listed in **Figure 2**.

Polarizability is a perennial theme in electronic materials research, driven in equal parts by applications and fundamental interest. We hypothesize that chalcogenides in crystal structures common to complex oxides may feature many highly-polarizable semiconductors. There are likely to be other chemical spaces worth searching, including continued exploration of halide perovskites and mixed-cation complex oxides.[53,56,60,61] These diverse research directions are linked by interest in how the dielectric response of crystals affects properties from minority carrier recombination rates to Cooper pairing. We look forward to continued exciting and productive research in this field, affecting applications from PV to telecommunications, and expanding our fundamental appreciation for electron and phonons in solids.

**Acknowledgments**

RJ acknowledges support from the National Science Foundation under contract 1751736, "CAREER: Fundamentals of Complex Chalcogenide Electronic Materials," and from the MIT Skoltech Program. JR acknowledges support from Air Force Office of Scientific Research (AFOSR grant number FA9550-16-1-0335) and Army Research Office (ARO grant number W911NF-19-1-0137).